\documentclass[12pt,english]{article}

\usepackage[T1]{fontenc}
\usepackage[latin9]{inputenc}
\usepackage[a4paper]{geometry}
\usepackage{float}
\usepackage{graphicx}
\usepackage[authoryear]{natbib}
\usepackage{subfig}
\usepackage{amsmath}

\makeatletter

\providecommand{\tabularnewline}{\\}

\newcommand{\lyxaddress}[1]{
\par {\raggedright #1
\vspace{1.4em}
\noindent\par}
}

\newcommand\blfootnote[1]{%
  \begingroup
  \renewcommand\thefootnote{}\footnote{#1}%
  \addtocounter{footnote}{-1}%
  \endgroup
}

\@ifundefined{showcaptionsetup}{}{%
 \PassOptionsToPackage{caption=false}{subfig}}
\usepackage{subfig}
\makeatother

\usepackage{babel}
\begin{document}

\title{Flow Past a Circular Cylinder on Curved Surfaces}

\author{Pankaj Jagad\textsuperscript{1,*}, Mamdouh S. Mohamed\textsuperscript{2},
Ravi Samtaney\textsuperscript{1}}

\date{}
\maketitle

\lyxaddress{\textsuperscript{1} Mechanical Engineering, Physical Science and
Engineering Division, King Abdullah University of Science and Technology,
Thuwal, Saudi Arabia}

\lyxaddress{\textsuperscript{2}Department of Mechanical Design and Production,
Faculty of Engineering, Cairo University, Giza, Egypt}

\blfootnote{
\textsuperscript{*}Corresponding author \\
Email addresses: pankaj.jagad@kaust.edu.sa (Pankaj Jagad), mamdouh.mohamed@kaust.edu.sa (Mamdouh S. Mohamed), ravi.samtaney@kaust.edu.sa (Ravi Samtaney) \\
}

\maketitle

\begin{abstract}
We investigate the dynamics of flows past a stationary circular cylinder
embedded on spherical and cylindrical surfaces at a fixed Reynolds number of 100. 
For flows on surfaces, it is convenient to express the Navier-Stokes equations  in exterior calculus notation. We employ the discrete exterior calculus (DEC) method (\cite{mohamed2016discrete}) to numerically solve the equations.  We explore the role played by curvature on dynamical quantities of interest: the drag coefficient $\left(C_{d}\right)$, the lift coefficient $\left(C_{l}\right)$, and the Strouhal number $\left(St\right)$. Our simulations indicate  that the effect of surface curvature on these quantities and the overall flow dynamics is somewhat insignificant thereby implying that the dynamics of the flow past a stationary circular cylinder exhibits universality, independent of the embedding surface geometry. 
\end{abstract}

\section{Introduction}

Research on the dynamics of flow past a circular cylinder
dates back to the century old discovery of K\'{a}rm\'{a}n vortex street (\cite{Karman1912}).
Some of the classic studies on this topic include works by K\'{a}rm\'{a}n, Taylor and others (\cite{Karman1912,taylor1915pressure,wieselsberger1921neuere,fage1929effects,thom1933flow}). Comprehensive reviews of the literature on this topic are available 
in \cite{niemann1990review,rao2015review}. It is evident from the literature that the dynamics of the flow past a cylinder embedded
on a flat surface have been extensively studied. One may pose the following question: what, if any, are the changes to the dynamics of flow past a circular cylinder embedded on a curved surface? So far a clear exploration of this question has not been undertaken. 
 This research problem is not only of academic interest, but also motivated by applications related to geophysical flows. 

In order to explore the research question expressed above, it is essential to have a robust numerical method with certain desirable conservation and mimetic properties. One such method is discrete exterior calculus (DEC).  
DEC is a mimetic method that retains, at the discrete level, many of the rules/identities of its continuous counterparts (\cite{hirani2003discrete,desbrun2005discrete}). This results in superior conservation properties for DEC discretization of physical problems. For example, many DEC-based discretizations of the incompressible Navier-Stokes equations are known to exactly conserve mass, vorticity and kinetic energy (\cite{mullen2009energy,mohamed2016discrete}).   This makes such discretizations  appropriate for investigating vortex dynamics and flows dominated by long-lived coherent structures. Moreover, DEC is coordinate independent and therefore convenient to investigate flows over arbitrary curved 
surfaces/manifolds. 

Here, we present a study of the dynamics of flows past a stationary
cylinder embedded on spherical and cylindrical surfaces. The 
DEC scheme developed by \cite{mohamed2016discrete} is used to perform the simulations. Surfaces comprising of different curvatures are employed, and the curvature effect on the flow dynamics is evaluated at a fixed Reynolds number ($Re=100$). 
A brief outline of this paper is as follows. The details of the physical setup, computational domain are in section \ref{sec:The-Computational-Domain}. The governing equations in the DEC framework are described in section \ref{sec:Numerical-Procedure}. The results are discussed in section \ref{sec:Results-and-Discussion}. The key outcomes of the study are finally summarized in section \ref{sec:Conclusion}.
 
\section{\label{sec:The-Computational-Domain}Physical Setup and Computational Domain}

\begin{figure} 
\begin{centering}
\subfloat[Spherical surface]{
\label{figure1a}
\begin{centering}
\includegraphics[scale=0.22]{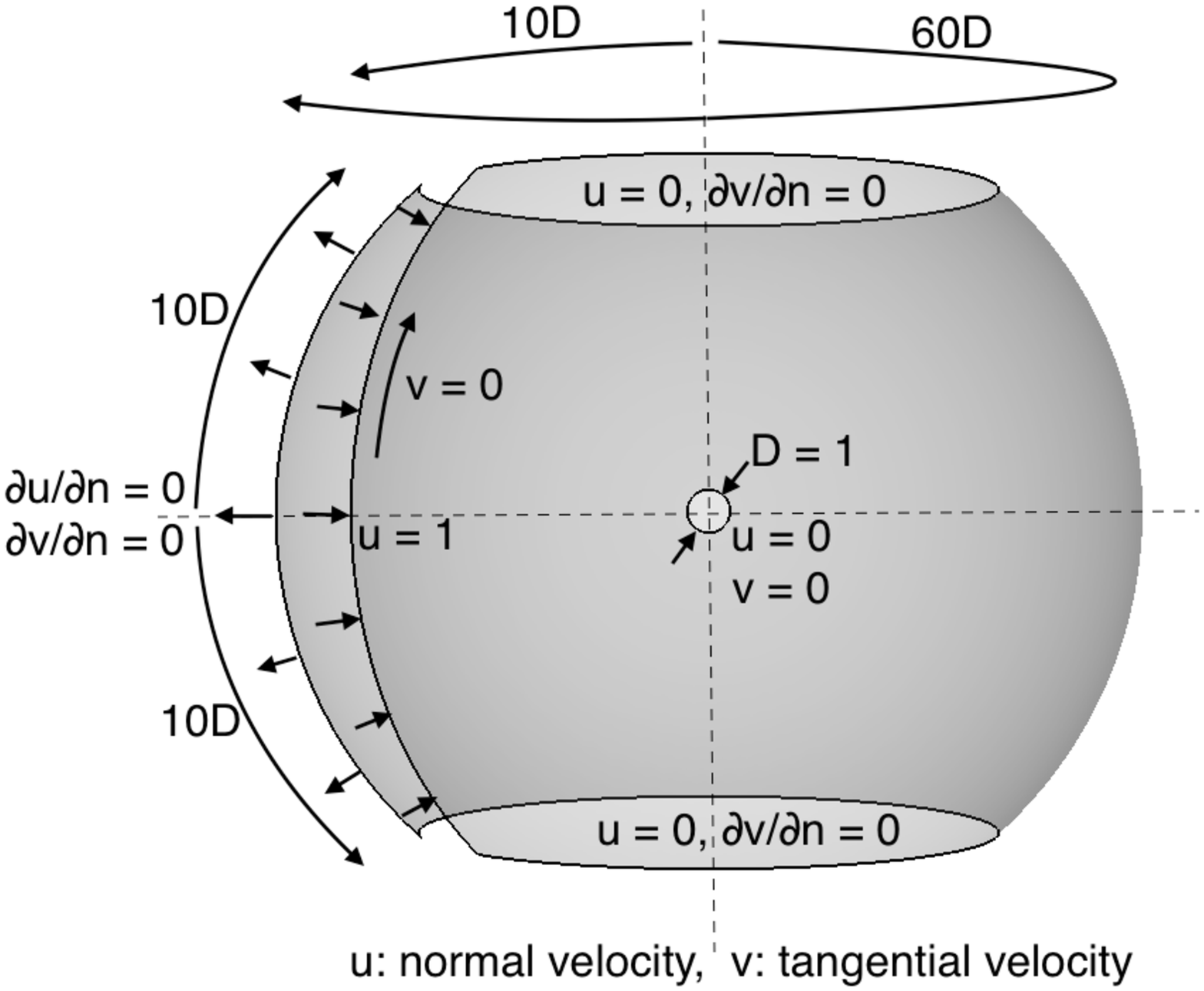}
\par\end{centering}
}\subfloat[\label{figure1b}Cylindrical surface]{
\begin{centering}
\includegraphics[scale=0.22]{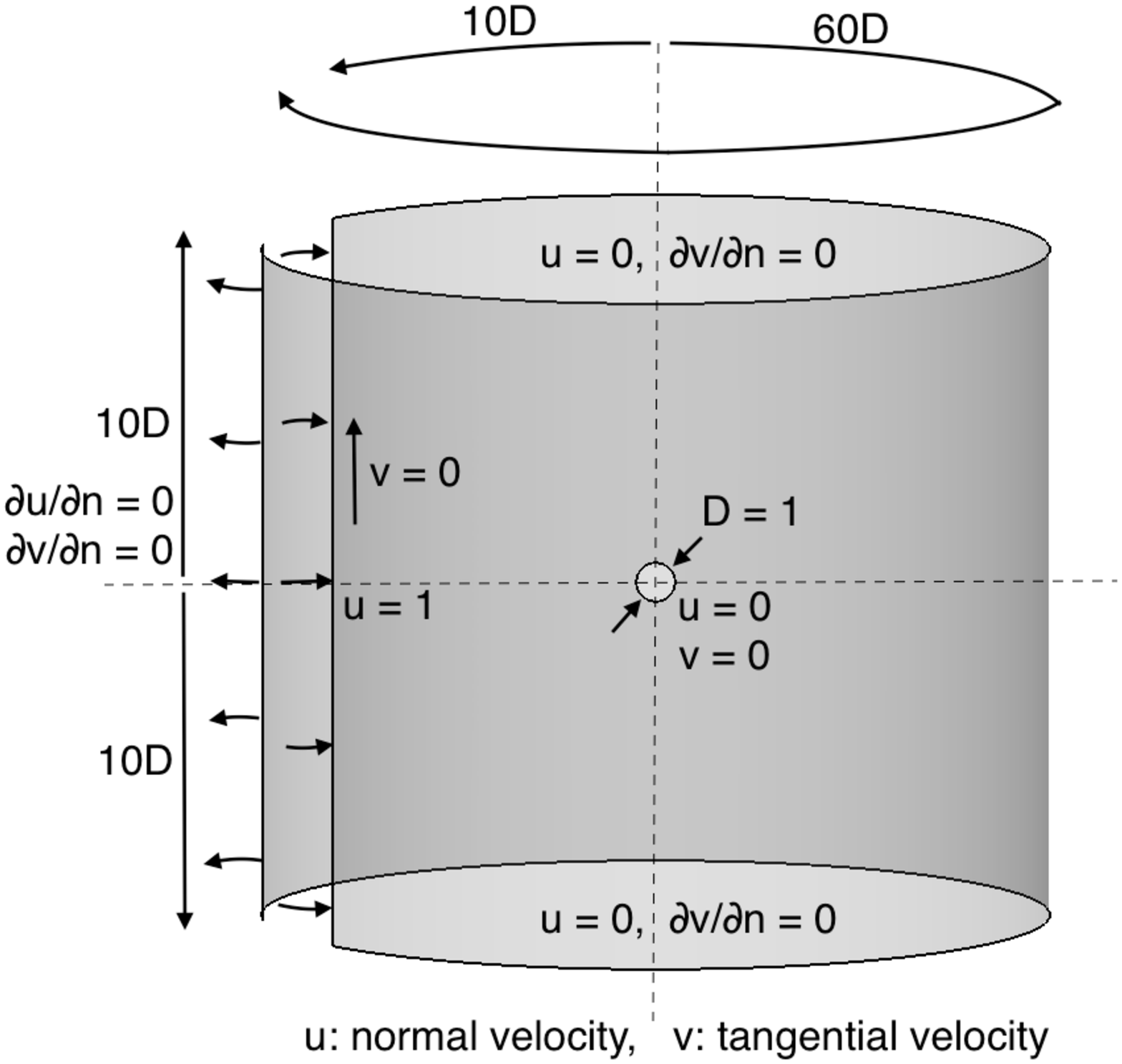}
\par\end{centering}
}\caption{\label{figure1} Schematic of the computational domain for both spherical and cylindrical surfaces showing the domain dimensions and the boundary conditions. Velocity components ($u,v$) indicate normal and tangential velocity components to the boundary.}
\par\end{centering}
\end{figure}

The schematic of the physical setup and computational domain are shown in figures \ref{figure1a} and \ref{figure1b} for the case of spherical and cylindrical surfaces, respectively. The circular cylinder contour embedded on the curved surfaces is generated by intersecting a cylinder of diameter D = 1 with the spherical and cylindrical surfaces. In terms of geodesic distances, the domain lengths upstream and downstream of the cylinder are 10D and 60D, respectively. The domain width on either side of the cylinder is 10D. These domain sizes are typical of computational investigations of flow past a cylinder on a flat surface. The implemented boundary conditions are shown in figure \ref{figure1} for each domain boundary, where $u$ and $v$ for a specific boundary represent the local normal and tangential velocity components. The radii of the embedding surfaces employed are reported in table \ref{table1}.

\section{\label{sec:Numerical-Procedure}Numerical Procedure}

The numerical scheme is discussed here only briefly, and the reader may refer to \cite{mohamed2016discrete} for more details. Unlike the stream function formulation employed in \cite{mohamed2016discrete}, the present simulations adopts the primitive variable formulation.
The Navier-Stokes equations, in exterior calculus notation, are expressed as 

\begin{equation}
\frac{\partial\mathbf{u}}{\partial t}-\mu\ast d\ast d\mathbf{u}-2\mu\kappa\mathbf{u}+\ast\left(\mathbf{u}\wedge\ast d\mathbf{u}\right)+dp^{d}=0,\label{eq:3_0}
\end{equation}
\begin{equation}
\ast d \ast \mathbf{u} = 0, 
\end{equation}

The reference for these equations is \cite{mohamed2016discrete}.  Note the presence of the term $2\mu\kappa\mathbf{u}$ : this is Gaussian curvature term which results from the divergence of the deformation tensor and the non-commutativity of the second covariant derivative in curved
spaces (\cite{nitschke2017discrete}). Here, $\mathbf{u}$, $\kappa$, $\mu$, $\ast$, $d$, $\wedge$, and $p^{d}$ denote the velocity 1-form, Gaussian curvature, dynamic viscosity, Hodge star, exterior derivative, wedge product, and dynamic pressure 0-form, respectively.

Considering the primal mesh, for the domain discretization, and the
corresponding dual mesh, equation (\ref{eq:3_0}) in DEC notation reads

\begin{equation}
\frac{U^{n+1}-U^{n}}{\triangle t}-\mu\ast_{1}d_{0}\ast_{0}^{-1}\left[\left[-d_{0}^{T}\right]U+d_{b}V\right]-2\mu\kappa U+\ast_{1}W_{V}\ast_{0}^{-1}\left[\left[-d_{0}^{T}\right]U+d_{b}V\right]+d_{1}^{T}P^{d}=0.\label{eq:1}
\end{equation}
Here, the choice of the velocity 1-form
$\mathbf{u}$ represents the velocity integration on the dual edges, except at the first position in the
wedge product where the choice has to be on the primal edge (this
choice represents tangential velocity 1-form $\mathbf{v}$) for consistency (\cite{mohamed2016discrete}). $U$, $V$, and $P^{d}$ are vectors representing the discrete
dual velocity 1-forms for all mesh dual edges, the primal velocity 1-forms for all mesh primal edges, and the dynamic pressure dual 0-forms for all mesh dual vertices, respectively. The discrete time interval is expressed as $\Delta t$. The $W_{V}$ matrix represents the discrete wedge product of the tangential
velocity 1-form with the 0-form \textasteriskcentered du (as in equation
(\ref{eq:3_0})) and contains the values of the tangential velocity
1-form. The operation $\left[-d_{0}^{T}\right]U$ evaluates the circulation of the velocity forms $\mathbf{u}$ along the dual 2-cells boundaries. The operation $d_{b}V$ complements the velocity circulation since a portion of the dual 2-cells boundaries may consists of primal edges, and accounts for the part that depends on the velocity 1-forms $\mathbf{v}$ on the primal boundary edges. 
The discrete operators $\ast_{1}$, $d_{0}$, $\ast_{0}^{-1}$, $\left[-d_{0}^{T}\right]$,
$d_{b}$, $W_{V}$, and $d_{1}^{T}$ are described in detail in Reference \cite{mohamed2016discrete}, and not repeated
here for the sake of brevity. Let $U^{\ast}$ be the vector containing mass flux primal 1-form $u^{\ast}$ for all mesh primal edges. We then have $U^{*}=-\ast_{1}^{-1}U$
and $U=\ast_{1}U^{\ast}$. After this substitution, equation (\ref{eq:1})
can be rewritten as 
\begin{equation}
\label{eq:2}
\begin{aligned}
\ast_{1}\left[\frac{\left(U^{*}\right)^{n+1}-\left(U^{*}\right)^{n}}{\triangle t}\right]&-\mu\ast_{1}d_{0}\ast_{0}^{-1}\left[\left[-d_{0}^{T}\right]\ast_{1}U^{*}+d_{b}V\right]-2\mu\kappa\ast_{1}U^{\ast}  \\ &+ \ast_{1}W_{V}\ast_{0}^{-1}\left[\left[-d_{0}^{T}\right]\ast_{1}U^{*}+d_{b}V\right]+d_{1}^{T}P^{d}=0.
\end{aligned} 
\end{equation}
Applying the Hodge star operator $\ast_{1}^{-1}$ to equation (\ref{eq:2}),
and considering the property $\ast_{1}^{-1}\ast_{1}=-1$,
we have 
\begin{equation}
\label{eq:3}
\begin{aligned}
-\frac{\left(U^{*}\right)^{n+1}-\left(U^{*}\right)^{n}}{\triangle t} &+\mu d_{0}\ast_{0}^{-1}\left[\left[-d_{0}^{T}\right]\ast_{1}U^{*}+d_{b}V\right]+2\mu\kappa U^{\ast}  \\ &-W_{V}\ast_{0}^{-1}\left[\left[-d_{0}^{T}\right]\ast_{1}U^{*}+d_{b}V\right]+\ast_{1}^{-1}d_{1}^{T}P^{d}=0.
\end{aligned} 
\end{equation}
In the above equation, the viscous and curvature terms are treated implicitly, while the convection term is treated explicitly. Hence, we write

\begin{equation}
\left[-\frac{1}{\triangle t}I+\mu d_{0}\ast_{0}^{-1}\left[-d_{0}^{T}\right]\ast_{1}+2\mu\kappa I\right]\left(U^{*}\right)^{n+1}+\ast_{1}^{-1}d_{1}^{T} \left(P^{d}\right)^{n+1}=F,
\label{eq:4}
\end{equation}
with
\begin{equation}
F=\frac{1}{\triangle t}\left(U^{*}\right)^{n}-\mu d_{0}\ast_{0}^{-1}d_{b}V^n+W_{V}^n\ast_{0}^{-1}\left[\left[-d_{0}^{T}\right]\ast_{1}\left(U^{*}\right)^{n}+d_{b}V^n\right].
\label{eq:5}
\end{equation}
The continuity equation is expressed as
\begin{equation}
\left[d_{1}\right]\left(U^{*}\right)^{n+1}+\left[0\right]\left(P^{d}\right)^{n+1}=0.
\label{eq:6}
\end{equation}

Equations (\ref{eq:4}) - (\ref{eq:6}) form a linear system of equations
with $U^{\ast}$ and $P^{d}$ at time $(n+1)$ to be the mass flux and dynamic pressure unknown degrees of freedom. This linear system is solved along with the boundary conditions depicted in figure \ref{figure1}. 

\begin{figure}
	\subfloat[\label{figure2a}Whole domain]{
		\begin{centering}
			\includegraphics[scale=0.3]{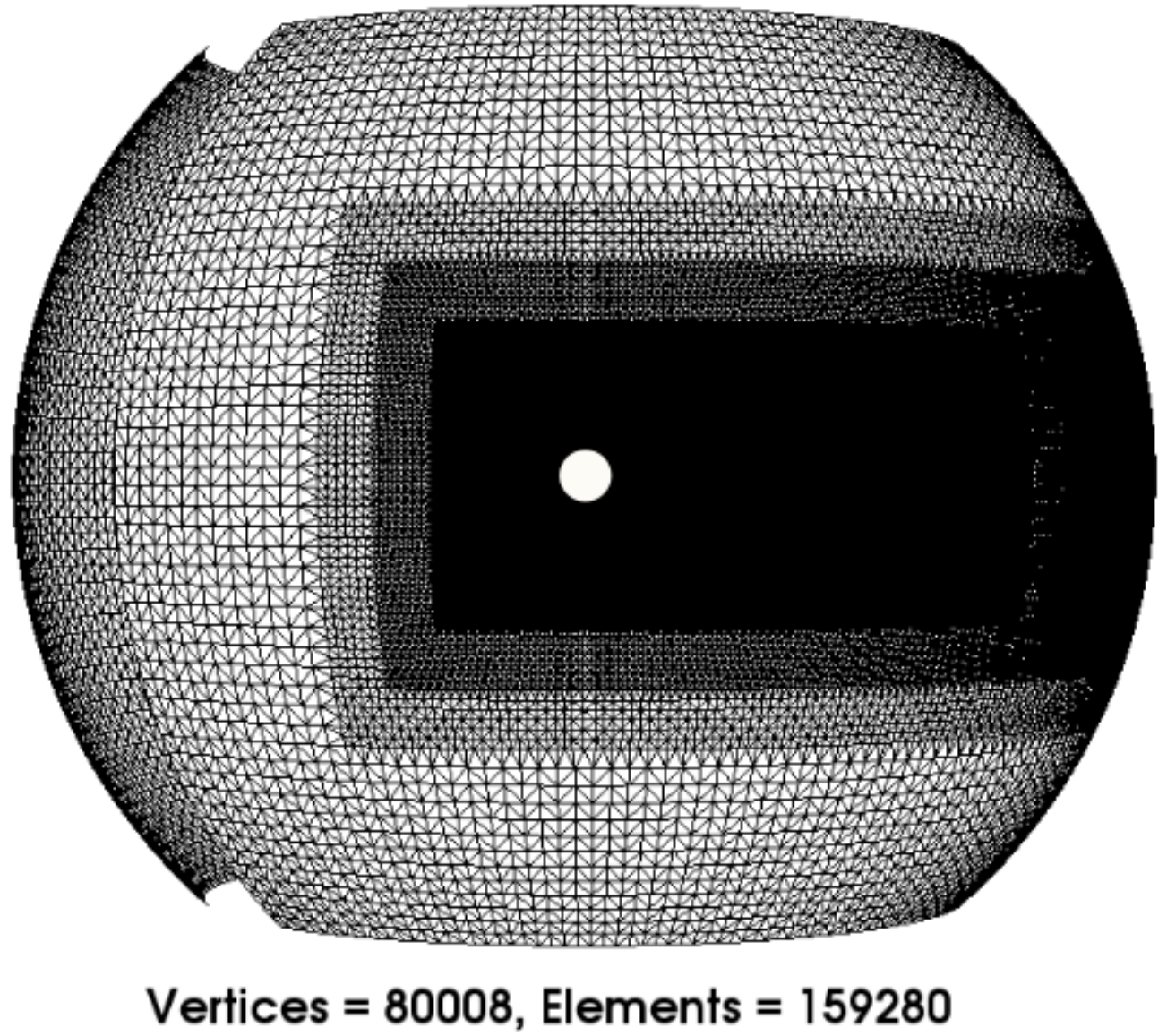}\includegraphics[scale=0.3]{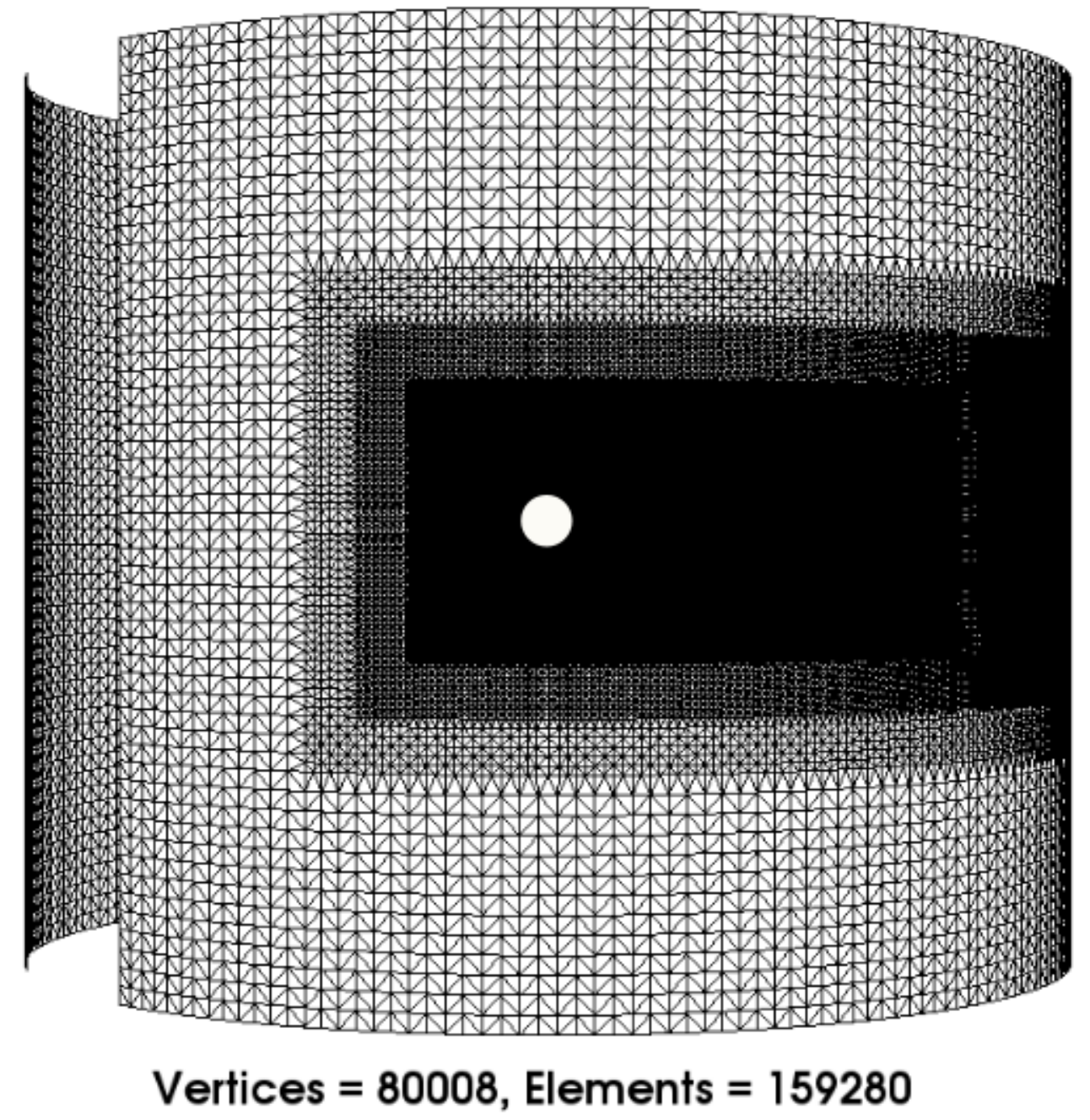}
			\par\end{centering}
	}
	\begin{centering}
		\subfloat[\label{figure2b}Around the cylinder]{\centering{}\includegraphics[scale=0.18]{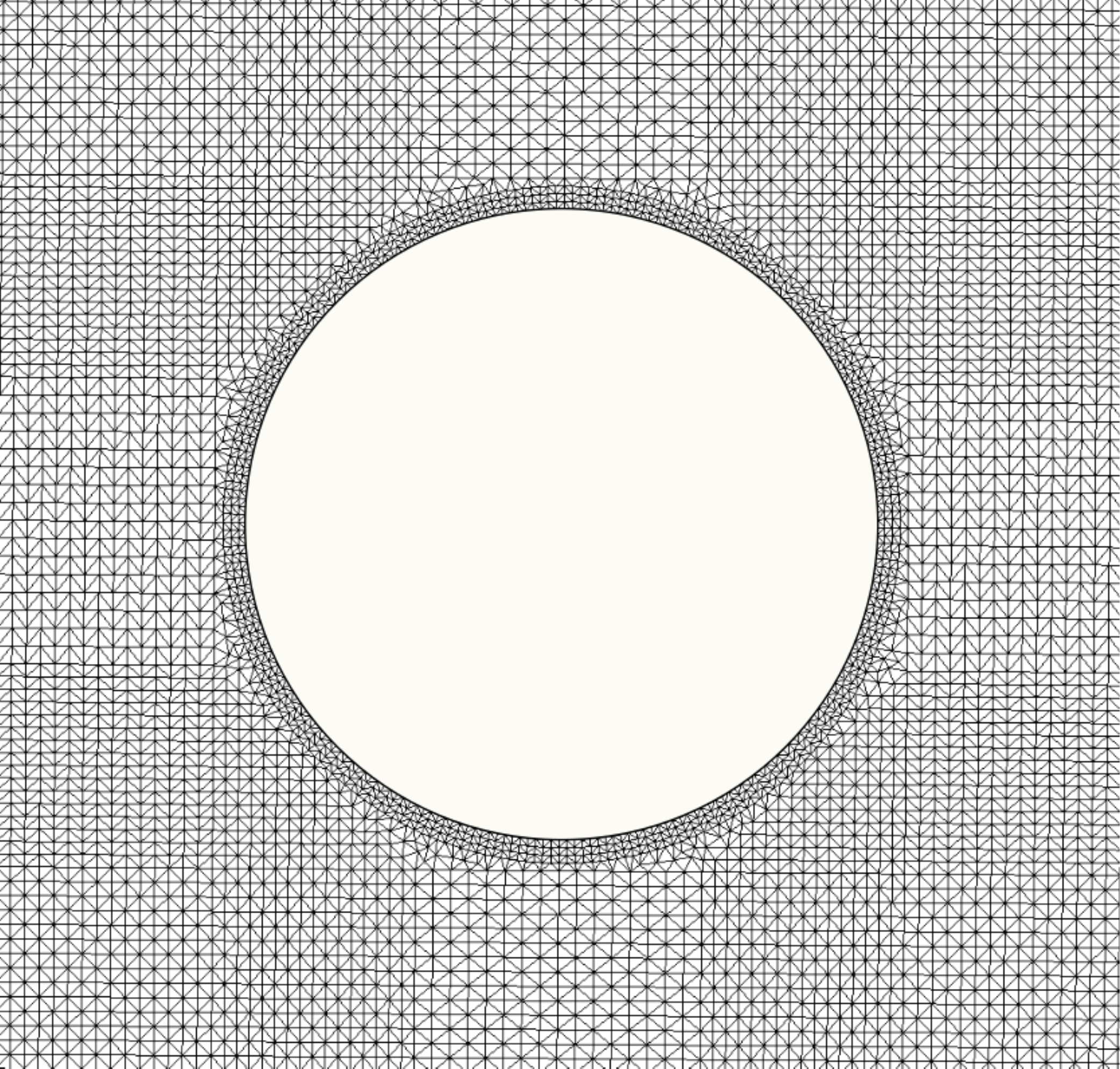}}\caption{\label{figure2}A representative computational mesh}
		\par\end{centering}
\end{figure}

The computational domain is discretized using a triangular mesh.
A representative computational mesh comprising of 80,008 nodes and
159,280 elements is shown in figure \ref{figure2}. Figure \ref{figure2a}
shows the mesh for the whole domain, whereas figure \ref{figure2b}
shows close up view around the cylinder. The nodes and elements for the meshes employed  are in the range 79,960 - 80,628, and 159,184-160,520, respectively. 

\section{\label{sec:Results-and-Discussion}Results and Discussion}
In this section, we present simulation results for flow past a cylinder with the 
configurations corresponding to (i) spherical surfaces and (ii) cylindrical surfaces, both with surface radii 
R = 12, 18, 24, 30, 36, and 72, and (iii) a flat surface as a reference. The Reynolds number ($Re=\rho u_{\infty}D/\mu$) is fixed at 100 for all the cases. Here, $\rho$, and $u_{\infty}$ denotes for the fluid density (taken to be unity), and the free stream velocity at the inlet, respectively.

The net force acting on the cylinder, due to pressure and viscous terms, is expressed as (\cite{Shi2004})
\begin{equation}
\vec{F}=\int_{S}-p\vec{n}dA+\mu\int_{S}\vec{\omega}\times\vec{n}dA
\end{equation}
where $p$ is the static pressure, $\vec{n}$ is the unit normal facing outward of the cylinder, $\vec{\omega}$ is the vorticity vector, $dA$ is an infinitesimal area (actually are lengths), and $S$ denotes for the total surface area of the cylinder. Since the boundary corresponding to the cylinder wall is discretized with $N$ mesh edges, the discrete expression of the force vector is 
\begin{equation}
\vec{F}=\sum_{k=1}^{N}-p_k \vec{n}_k l_k+\mu\sum_{k=1}^{N} l_k \vec{\omega}_k \times \vec{n}_k
\end{equation}
where the summation is over the cylinder boundary edges. Here, we consider the global $(x,y,z)$ coordinate system with the cylinder axis intersection with the surface as the origin, the cylinder axis along the z-direction, and the x and y  directions as shown in figure \ref{figure1}. The pressure $p_k$ is the static pressure in the neighbor triangle, $\vec{n}_k$ is the unit vector normal to the edge (projected on the x-y plane), and $l_k$ is the edge length. For the case of 2D surfaces, the vorticity is calculated as a scalar on the primal mesh nodes (\cite{mohamed2016discrete}). The vector $\vec{\omega}_k=\vec{n}_{k,s}|\omega_k|$, with $\vec{n}_{k,s}$ denoting the unit normal facing outward to the surface at the edge midpoint. The vorticity magnitude $|\omega_k|$ is calculated by averaging the vorticity on both nodes of the edge. The drag force $F_{d}$ and the lift force $F_{l}$ are respectively the $x$ and $y$ components of the force vector.

The drag coefficient is calculated using the expression $C_{d}=F_{x}/\left(\frac{1}{2}\rho u_{\infty}^{2}D\right)$, where $F_{x}$ denotes for the x-component of the net force on the cylinder. To calculate the lift
coefficient, the expression $C_{l}=F_{y}/\left(\frac{1}{2}\rho u_{\infty}^{2}D\right)$
is used, where $F_{y}$ is the y-component of the net force on the
cylinder. The Strouhal number is
$St=fD/u_{\infty}$, where $f$ is the vortex shedding frequency.

\begin{figure}
	\begin{centering}
		\subfloat[\label{figure3_0a}Spherical surface]{
			\begin{centering}
				\includegraphics[scale=0.16]{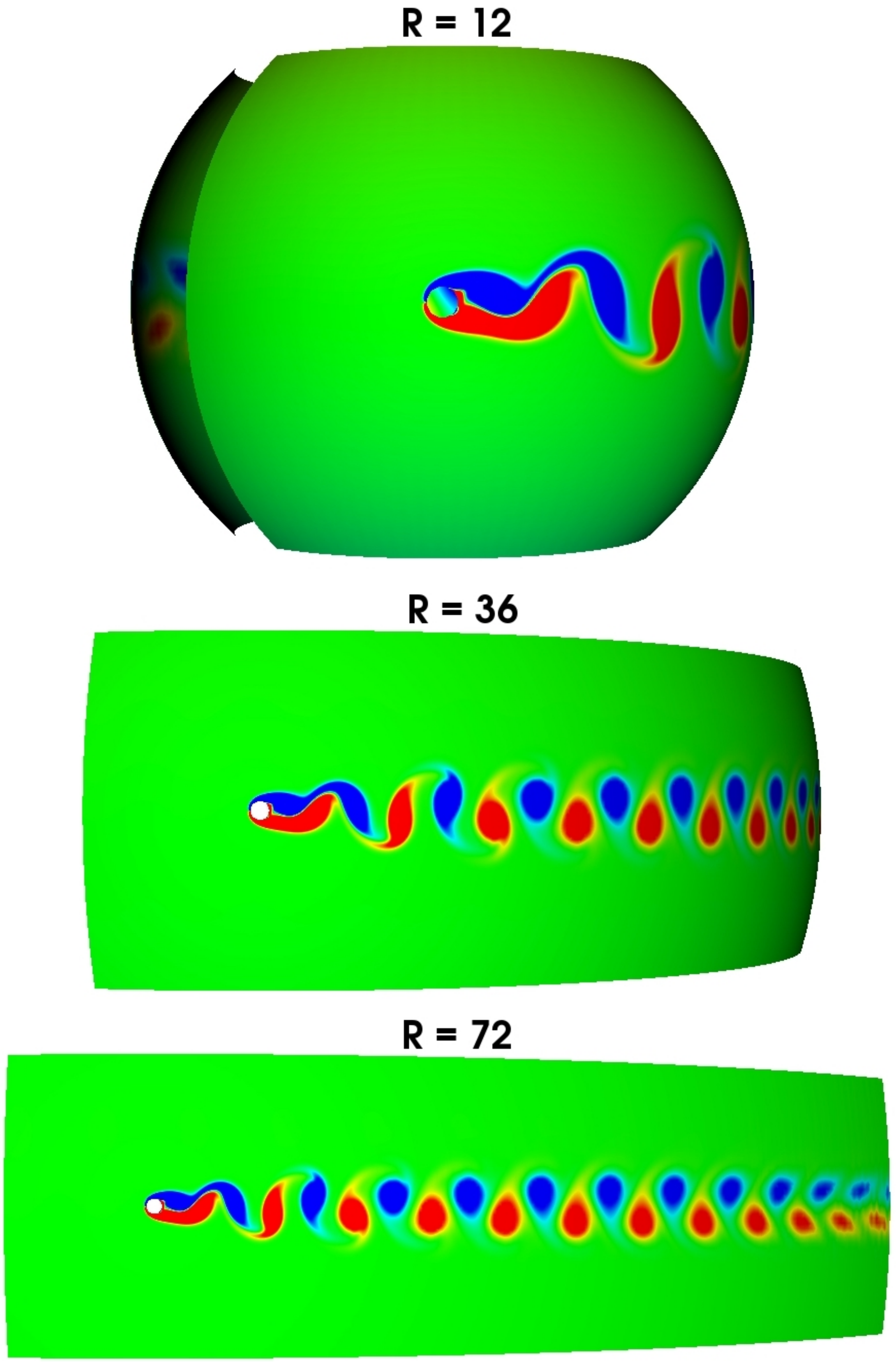}
				\par\end{centering}
		}\subfloat[\label{figure3_0b}Cylindrical surface]{
			\centering{}\includegraphics[scale=0.16]{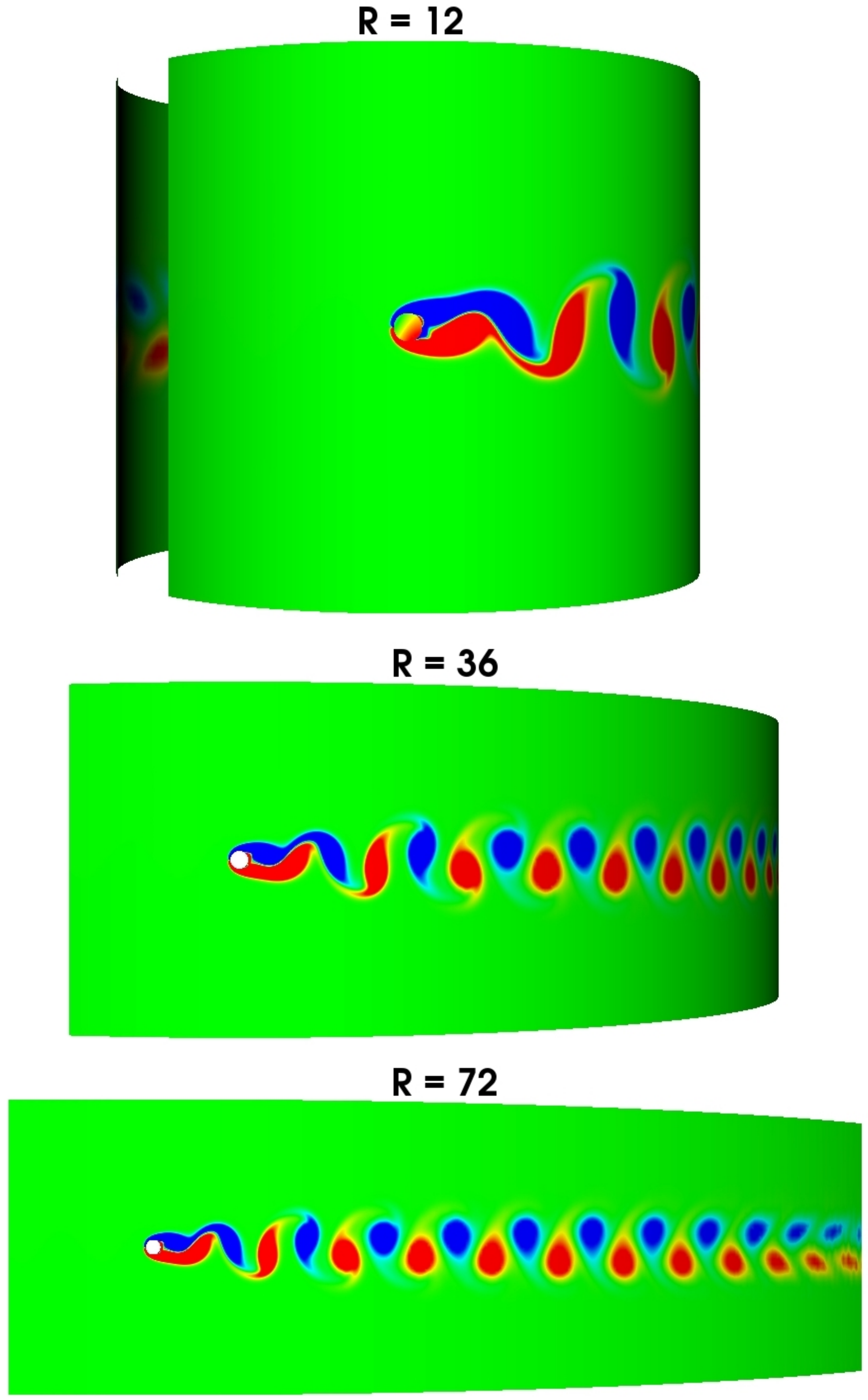}}
		\par\end{centering}
	\caption{\label{figure3_0}Vortex streets. R denotes radius of the embedding surface.}
\end{figure}

\begin{figure}
	\begin{centering}
		\subfloat[\label{figure3a}Spherical surface]{\begin{centering}
				\includegraphics[scale=0.16]{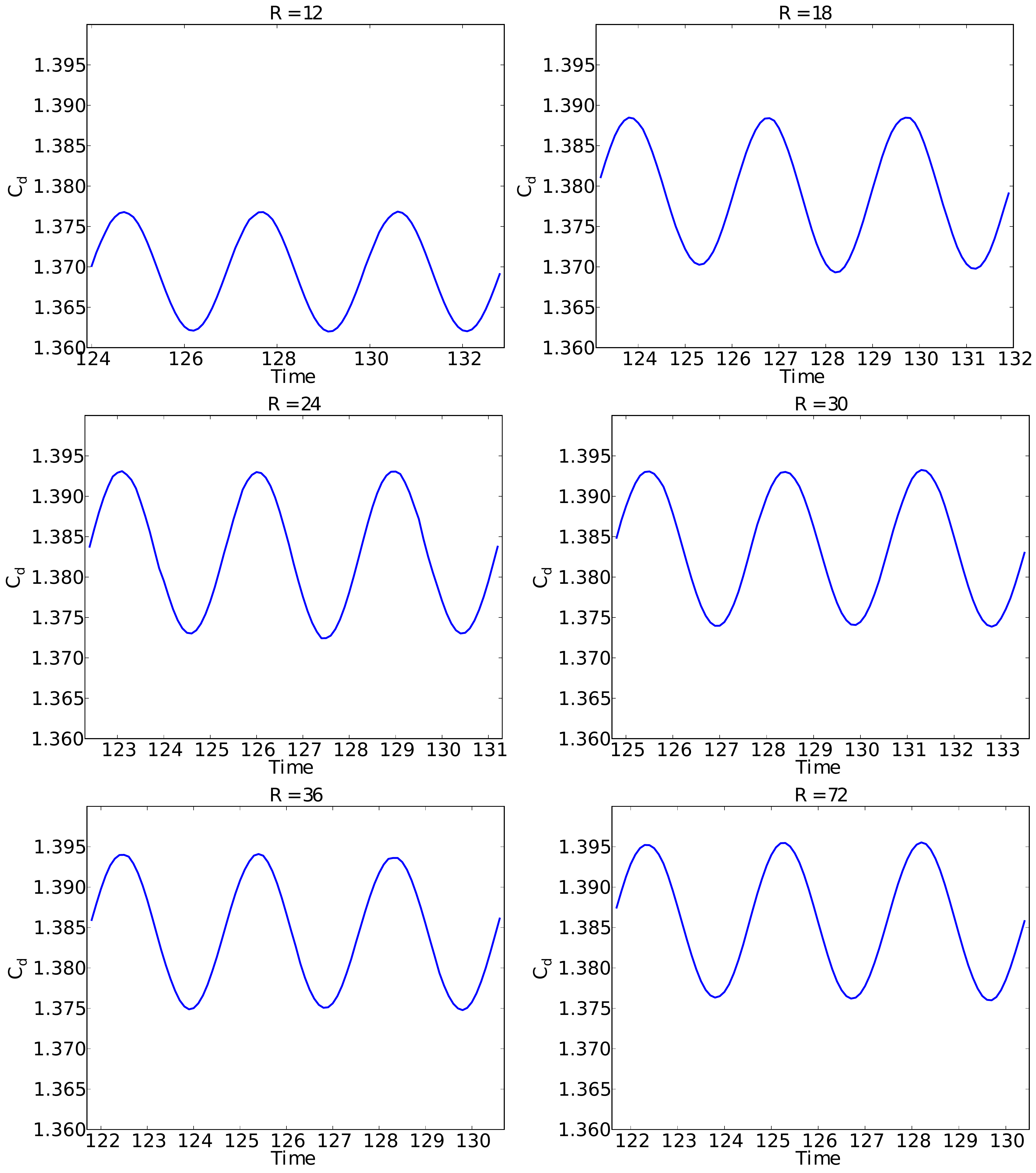}
				\par\end{centering}
		}\subfloat[\label{figure3b}Cylindrical surface]{
			\begin{centering}
				\includegraphics[scale=0.16]{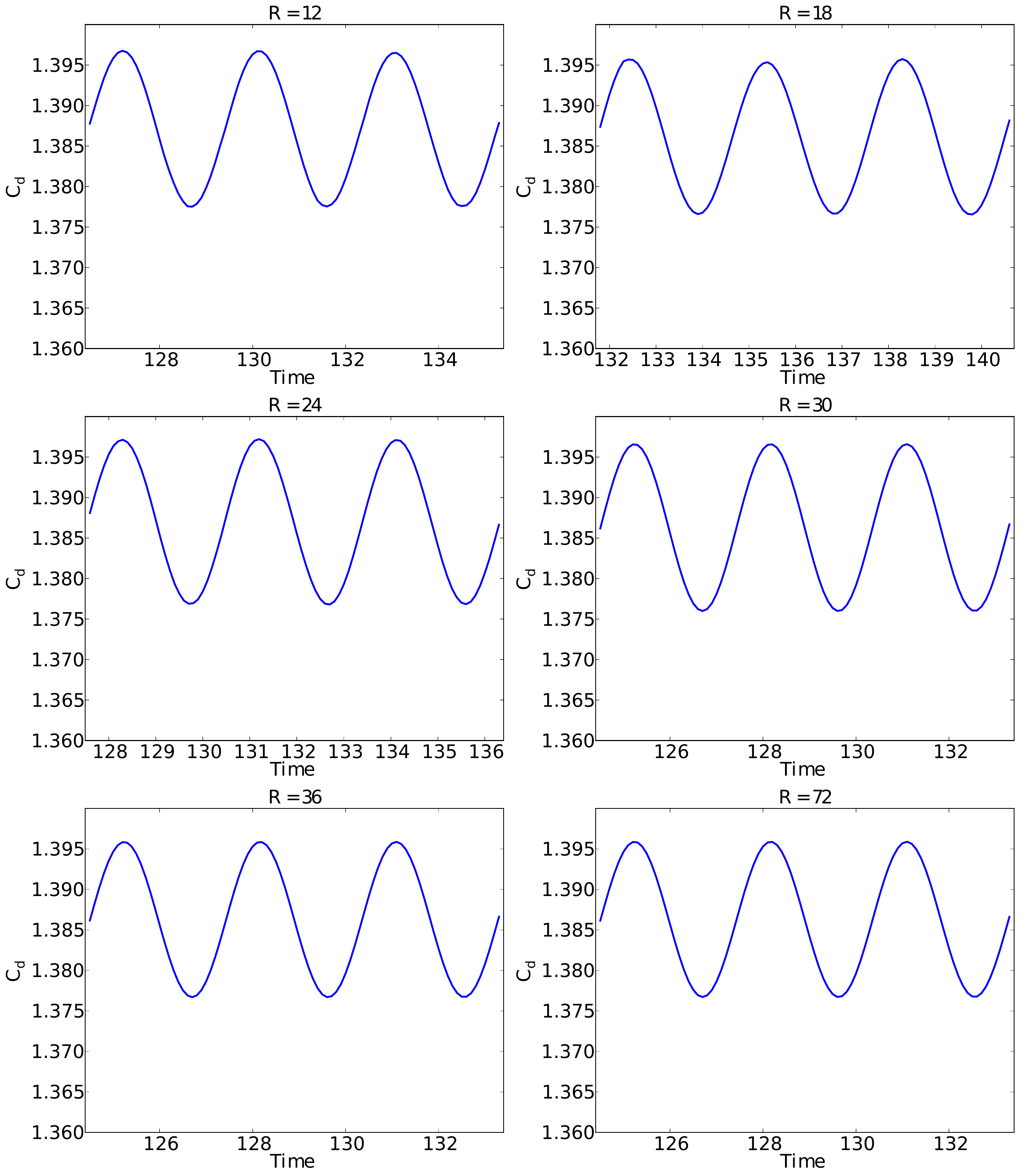}
				\par\end{centering}
		}
		\par\end{centering}
	\begin{centering}
		\subfloat[\label{figure3c}Flat surface]{\begin{centering}
				\includegraphics[scale=0.16]{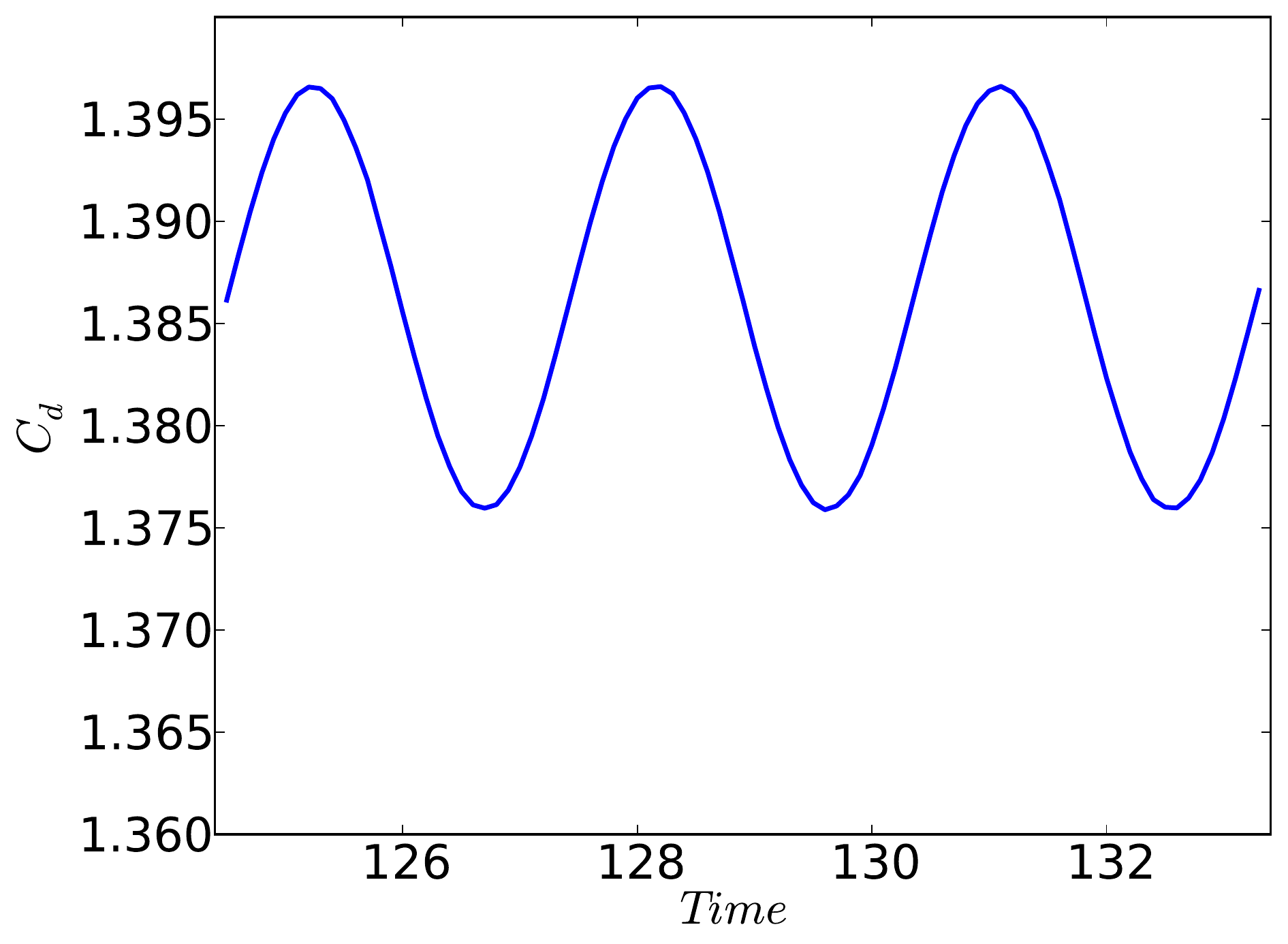}
				\par\end{centering}
		}\caption{Drag coefficient as a function of time. R denotes radius of the embedding surface }
		\par\end{centering}
\end{figure}

Figure \ref{figure3_0} shows the K\'{a}rm\'{a}n vortex street in the wake of each cylinder for different radii of the embedding cylindrical or spherical surface. The vortices in magnitude and spacing are similar to those for the flow past a cylinder on a flat surface. The plots of the drag coefficient, $C_d$ as a function of time, for a finite time duration
after the periodically steady vortex shedding commences, are shown in
figures \ref{figure3a}, \ref{figure3b}, and \ref{figure3c} for
spherical, cylindrical and flat surfaces, respectively, and the corresponding
mean values are reported in table \ref{table1}. The mean drag coefficient
for the flat surface is  $1.386$, which is within the range of values reported in the literature (i.e., 1.38 (\cite{russell2003cartesian}), 1.37 (\cite{le2006immersed})). For the case of spherical surface, the mean drag coefficient is $1.386$ for $R=72$. The mean drag coefficient decreases slightly with increasing curvature of the surface, and for $R=12$ (the highest curvature case) it is computed to be $1.369$.
The change in $C_d$ with curvature is about $1\%$ and thus very weakly dependent (or perhaps even independent) of the embedding surface  curvature, and the mean drag coefficient almost
coincides with that for the flat surface. For the case of embedding cylindrical
surface, the mean drag coefficient is constant at $1.386$ coinciding
with that for the flat surface. Relative changes in magnitude of $C_d$ of about $1\%$ or a few percent are not unexpected when the mesh resolution changes; or a different quadrature used to compute the drag coefficient by integrating the forces on the surface of the cylinder. One may reasonably conclude that the effect of the embedding surface curvature on the drag coefficient is insignificant.

\begin{figure}
	\begin{centering}
		\subfloat[\label{figure4a}Spherical surface]{
			\begin{centering}
				\includegraphics[scale=0.16]{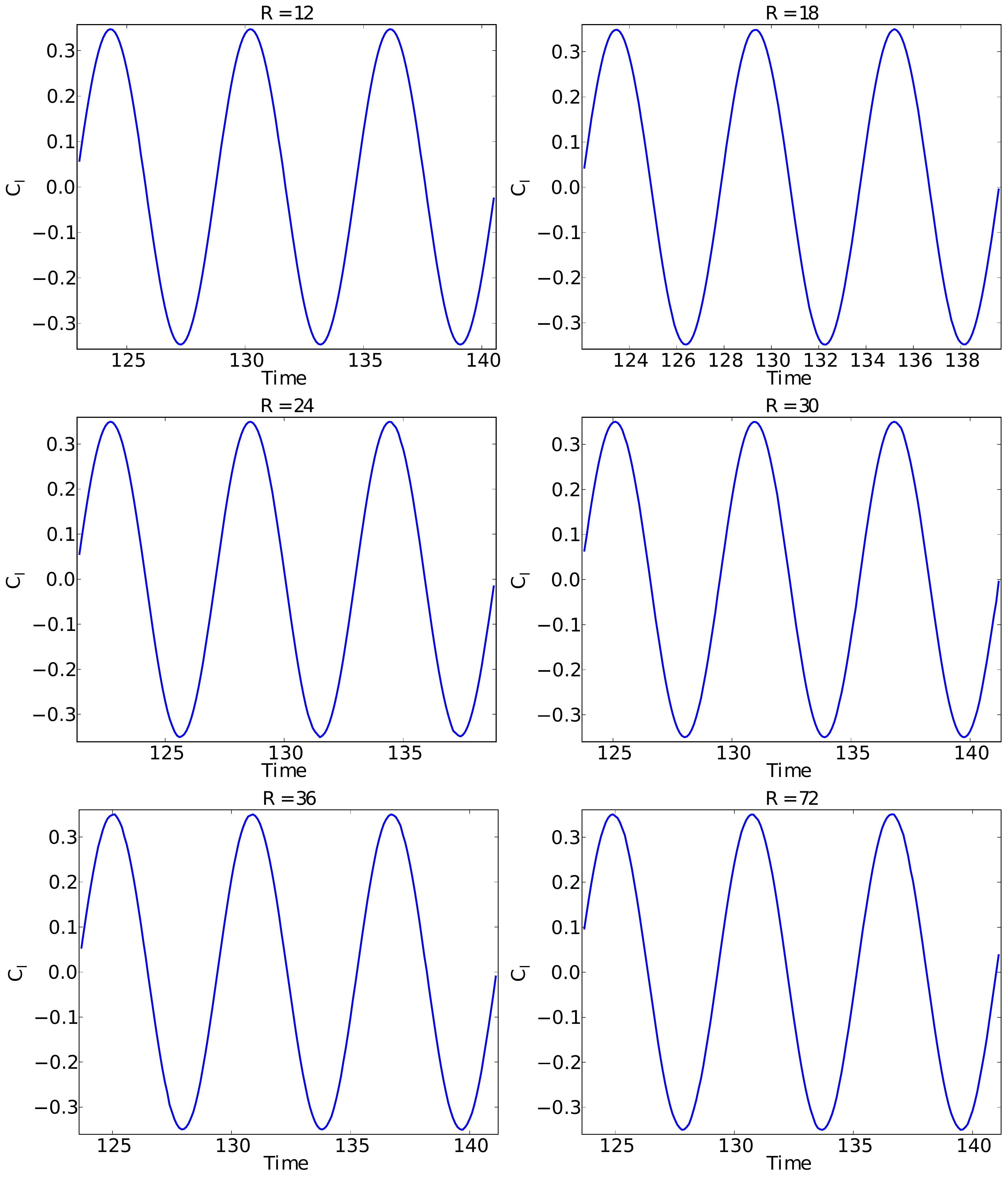}
				\par\end{centering}
		}\subfloat[\label{figure4b}Cylindrical surface]{
			\centering{}\includegraphics[scale=0.16]{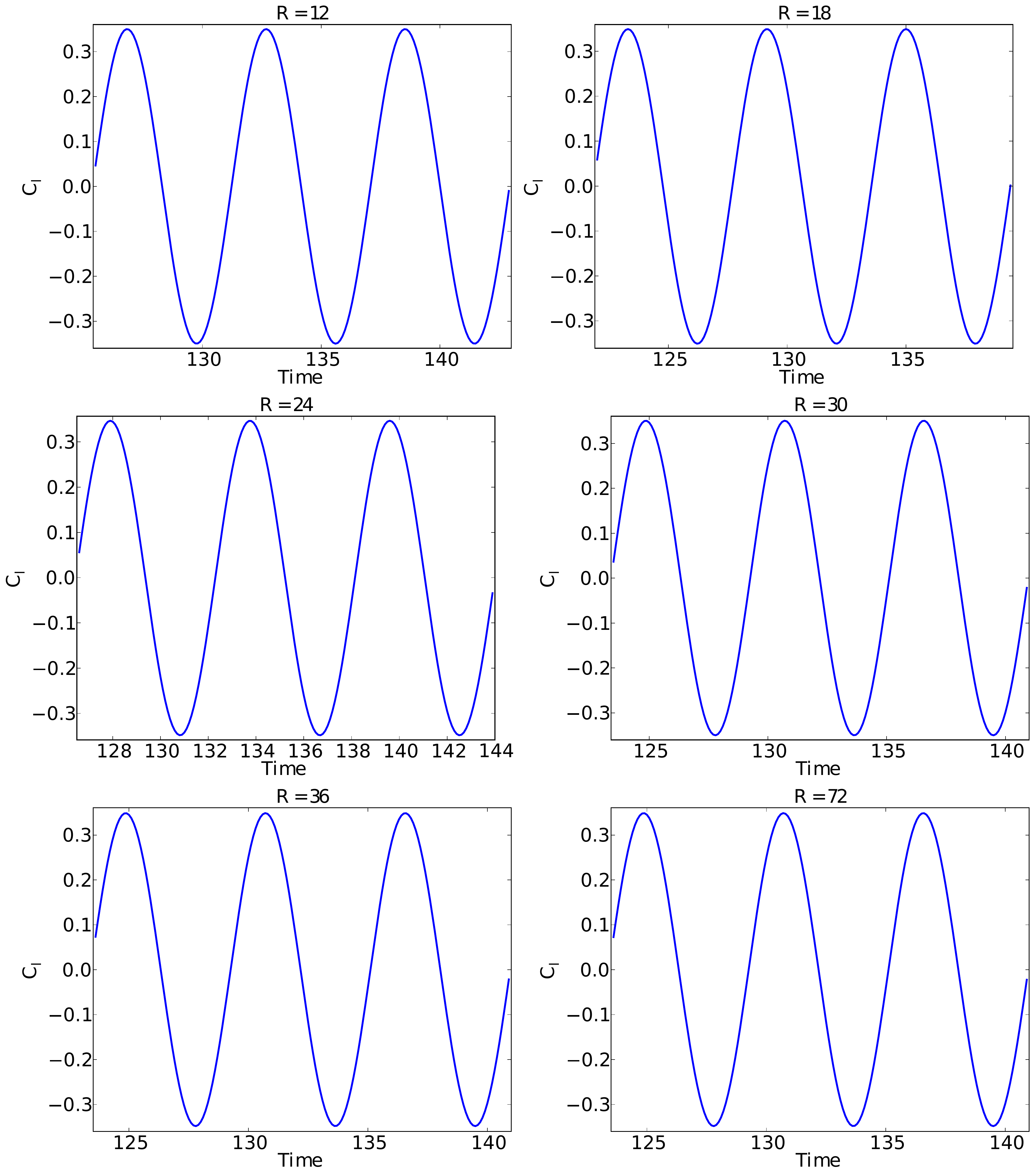}}
		\par\end{centering}
	\centering{}\subfloat[\label{figure4c}Flat surface]{\begin{centering}
			\includegraphics[scale=0.16]{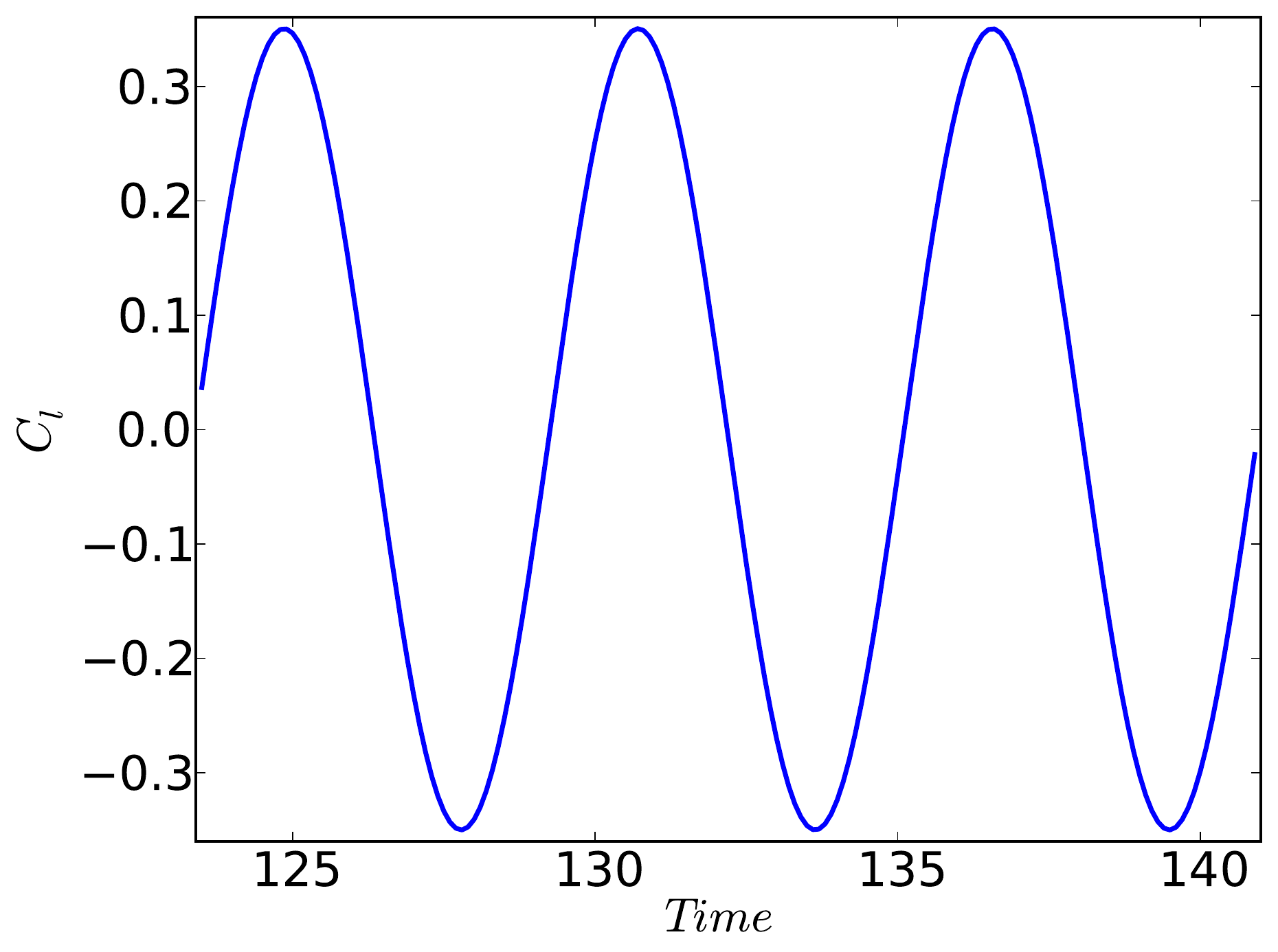}
			\par\end{centering}
	}\caption{Lift coefficient as a function of time.}
\end{figure}

Figures \ref{figure4a}, \ref{figure4b}, and \ref{figure4c} show
the lift coefficient as a function of time for the cases
of spherical, cylindrical and flat surfaces, respectively. The RMS
values of the lift coefficient are reported in table \ref{table1}.
The RMS lift coefficient value is $0.248$ for the flat surface, which is within the range of values reported in the literature. It is computed to to be $0.248$ for the spherical surface of $R=72$ and $ 0.246$ for $R=12$, i.e. a negligible decrease (less than $1\%$) in the value with changes in curvature of the embedding surface. For the cylindrical surface, all RMS values are constant at $0.248$ coinciding with that for the flat surface. These results suggest insignificant effect of the surface curvature on the lift coefficient RMS value. It is worth noting that there exists a z-component of the net force on the cylinder due to the viscous term, but this is negligibly small for both the spherical and cylindrical embedding surfaces. 

The values of shedding frequency or Strouhal number are reported in table \ref{table1}.
For the flat surface, the Strouhal number is computed as $0.171$ which is
within the range of values reported in the literature. For the spherical surface,
the Strouhal number is $0.171$ for $R=72$. It changes negligibly with the
radius, and for $R=12$ it is found to be $0.170$. Thus, it almost
coincides with that for the flat surface. The Strouhal number is almost
constant at $0.171$ irrespective of the radius for the cylindrical
surfaces, which coincides with that for the flat surface. These results  suggest the negligible effect of the surface curvature on the Strouhal number.

\begin{table}

\centering{}%
\begin{tabular}{|c|c|c|c|c|}
\hline 
Surface type & Radius of the surface & $C_{d}$ (mean) & $C_{l}$ (RMS) & $St$\tabularnewline
\hline 
\hline 
Spherical surface & 12 & 1.369 & 0.246  & 0.17 \tabularnewline
\cline{2-5} 
 & 18 & 1.379  & 0.247  & 0.17 \tabularnewline
\cline{2-5} 
 & 24 & 1.383  & 0.247  & 0.171 \tabularnewline
\cline{2-5} 
 & 30 & 1.384  & 0.248  & 0.171 \tabularnewline
\cline{2-5} 
 & 36 & 1.384  & 0.248  & 0.171 \tabularnewline
\cline{2-5} 
 & 72 & 1.386 & 0.248 & 0.171\tabularnewline
\hline 
Cylindrical surface & 12 & 1.386  & 0.248  & 0.171\tabularnewline
\cline{2-5} 
 & 18 & 1.386 & 0.248 & 0.171\tabularnewline
\cline{2-5} 
 & 24 & 1.386 & 0.248 & 0.171\tabularnewline
\cline{2-5} 
 & 30 & 1.386 & 0.248 & 0.171\tabularnewline
\cline{2-5} 
 & 36 & 1.386 & 0.248 & 0.171\tabularnewline
\cline{2-5} 
 & 72 & 1.386 & 0.248 & 0.171\tabularnewline
\hline 
Flat surface  & Present study & 1.386 & 0.248 & 0.171\tabularnewline
\cline{2-5} 
 & \cite{russell2003cartesian} & 1.38 & - & 0.172\tabularnewline
\cline{2-5} 
 & \cite{le2006immersed} & 1.37 & - & 0.160\tabularnewline
\cline{2-5} 
 & \cite{Park1998} & 1.33 & 0.3321 & 0.165\tabularnewline
\cline{2-5} 
 & \cite{Mittal2005} & 1.322 & 0.226 & 0.1644\tabularnewline
\cline{2-5} 
 & \cite{Stalberg2006} & 1.32 & 0.233 & 0.166\tabularnewline
\cline{2-5} 
 & \cite{Posdziech2001} & 1.325 & 0.228 & 0.1644\tabularnewline
\cline{2-5} 
 & \cite{Li2009} & 1.336 & - & 0.164\tabularnewline
\cline{2-5} 
 & \cite{Qu2013} & 1.319 & 0.225 & 0.1648\tabularnewline
\hline 
\end{tabular}\caption{\label{table1}Values of the drag coefficient, the lift coefficient and the Strouhal number. }
\end{table}

\begin{figure}
	\begin{centering}
		\subfloat[\label{figure5a}Pressure coefficient]{
			\begin{centering}
				\includegraphics[scale=0.22]{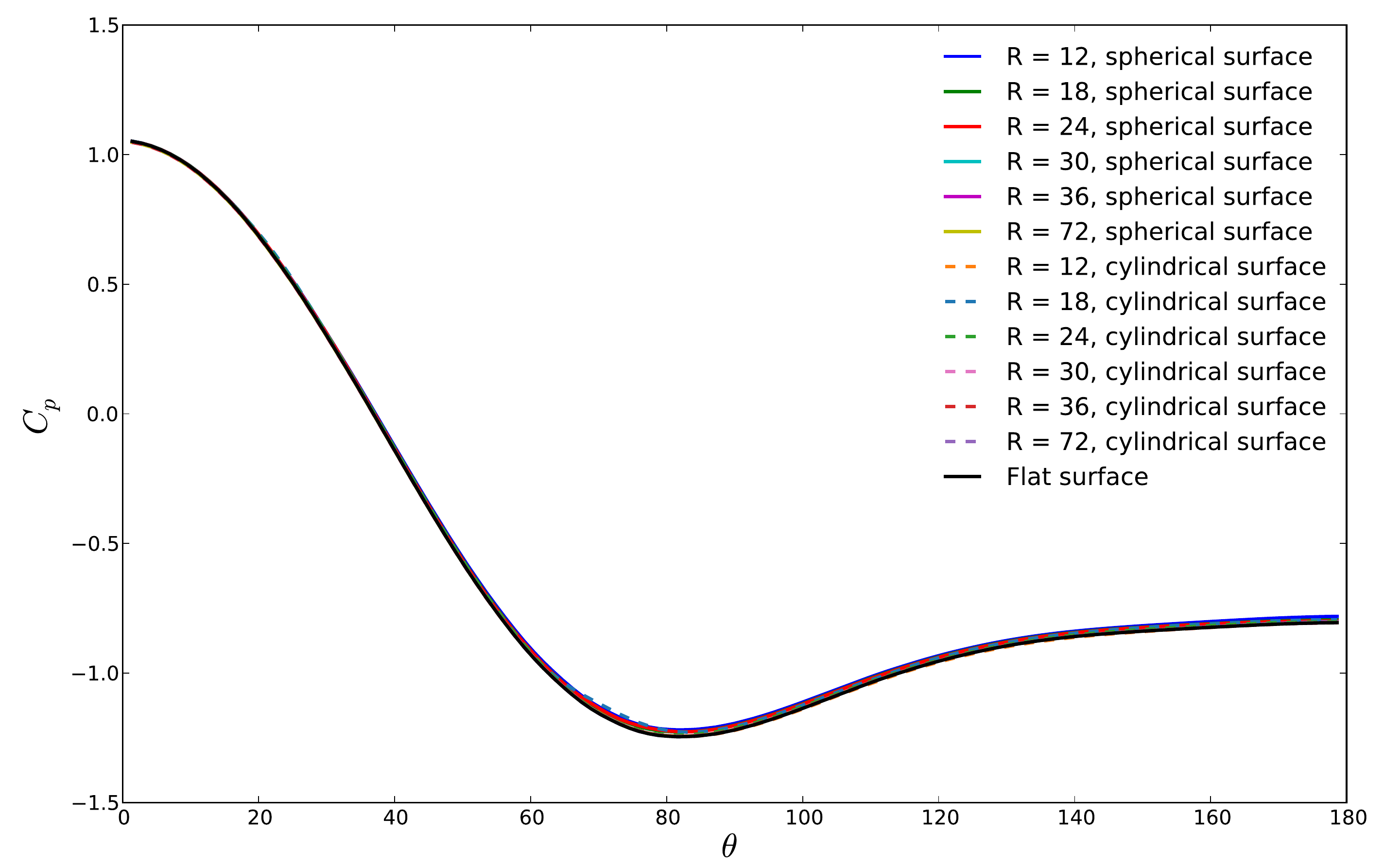}
				\par\end{centering}
		}\subfloat[\label{figure5b}Vorticity magnitude]{
			\centering{}\includegraphics[scale=0.22]{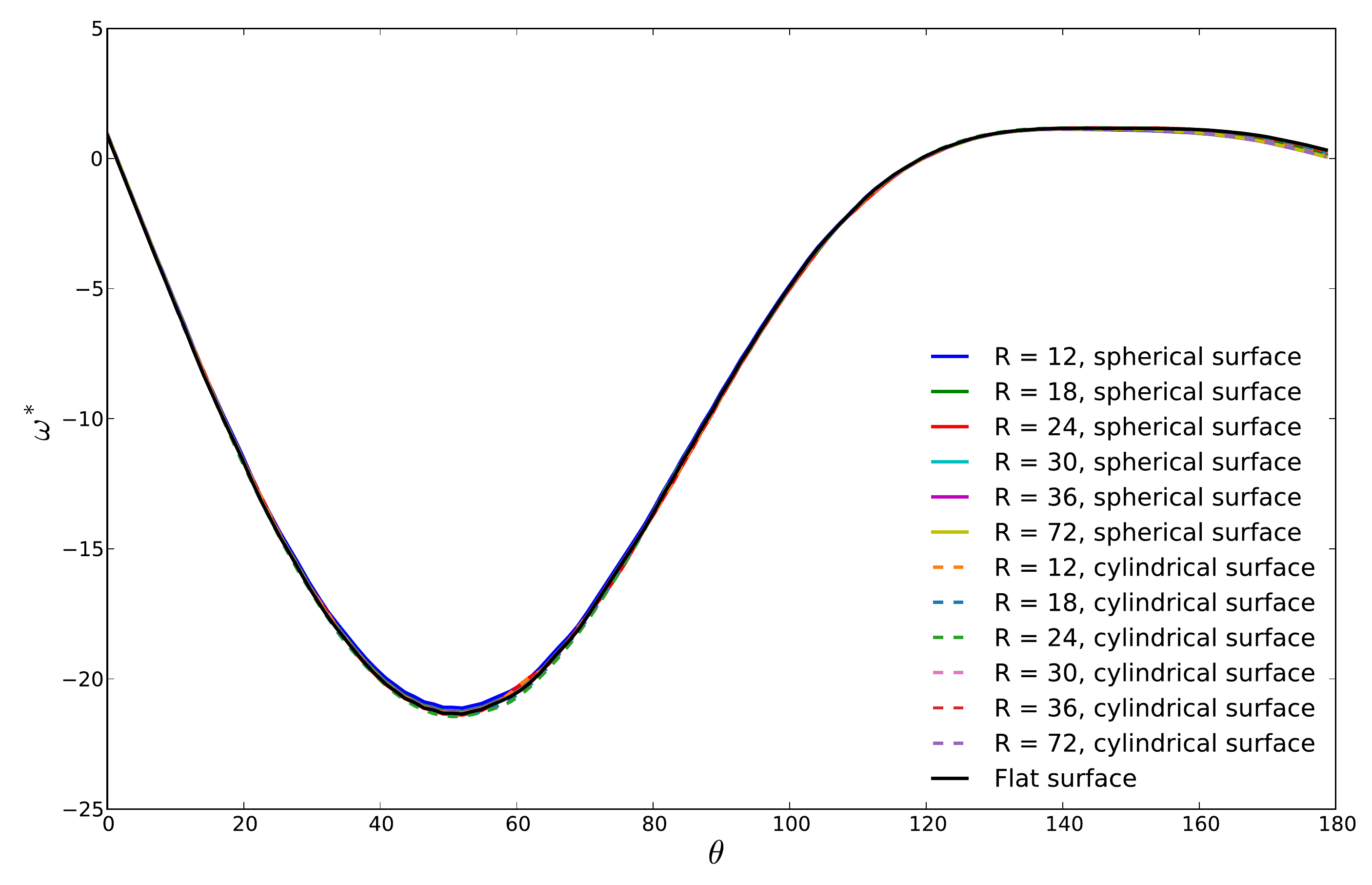}}
		\par\end{centering}
	\caption{Pressure coefficient and vorticity magnitude as a function of the angular
		position on the cylinder surface.}
\end{figure}

The pressure coefficient plots $C_{p}=\left(p-p_{\infty}\right)/\left(\frac{1}{2}\rho u_{\infty}^{2}\right)$ as a function of the angular position on the cylinder surface, for all surfaces of different radii as well as the flat surface, are
shown in figure \ref{figure5a}, where 
$p_{\infty}$ denotes the free stream pressure. The plots 
overlap, which further supports the negligible effect of the surface curvature on the flow dynamics. Similarly, 
we note the absence of any significant dependence on curvature in the plots of vorticity magnitude $\omega^{\ast}=\omega_{n}D/u_{\infty}$ as a function of the angular position on the cylinder surface as shown in figure \ref{figure5b}, where $\omega_{n}$ is the vorticity in the surface normal direction (considered as a scalar for calculations on surfaces). 

The aforementioned results all suggest that the dynamics of the flow past
a stationary circular cylinder embedded on a curved surface is not
significantly affected by the surface curvature, and is almost identical
to that for a flat surface. In other words, the dynamics of flow past a circular cylinder exhibits universality.
For a given embedded cylinder, in addition to the geometry of the embedding surface,  there are two other factors
that can affect the flow dynamics around the cylinder on a curved surface, the
Gaussian curvature term $\left(2\mu\kappa U\right)$ in equation (\ref{eq:1})
and the departure of the intersection contour between the cylinder
and the surface from being a planar circle. In order to quantify the relative magnitude of  curvature term influence in the momentum equation, we compared its magnitude with the viscous term magnitude (both terms depend on the fluid viscosity or Reynolds number). For the case of spherical embedding surface, we found that the viscous term  ($\mu\ast_{1}d_{0}\ast_{0}^{-1}\left[\left[-d_{0}^{T}\right]U+d_{b}V\right]$) dominates in the region around the cylinder, being almost eight orders of magnitude larger than the curvature term. Therefore, for the simulations considered in the present study, the curvature term in equation (\ref{eq:1}) has insignificant effect on the flow dynamics around the cylinder. For a cylindrical surface on the other hand, the Gaussian curvature, and therefore the curvature term, is identically equal to zero and does not affect the flow dynamics. Let us now consider the aspect of the  domain geometry, i.e., the intersection contour between the cylinder and the embedding surface. For a spherical embedding surface, the intersection is identical to a planar circle. The geometry in the region around the cylinder is therefore not significantly different from that for a flat surface, and does not considerably affect the flow dynamics.  On the other hand, for an embedding cylindrical surface, the intersection contour with the cylinder is not planar. However, for the cases considered during the present study, the differences between such intersection contour and a plane circle remains relatively small. This can be quantified by evaluating the maximum variation of the z-coordinate along the intersection contour and also the change in the intersection contour length in comparison to a plane circle, as stated in table \ref{table2}. The reported values are evidently insignificant. The inconsiderable impact of the curvature term in equation (\ref{eq:1}) along with the trivial domain geometry changes ultimately resulted in the observed similarity in the flow past a cylinder dynamics on curved surfaces, in comparison to that on a flat surface.

\begin{table}
\centering{}%
\begin{tabular}{|c|c|c|}
\hline 
Radius of cylindrical surface  & Maximum z -- variation  & \% change in the contour \tabularnewline
 & along the contour &  length when compared\tabularnewline
 &  & to that for a flat surface\tabularnewline
\hline 
\hline 
R = 12  & 0.01040 & 0.02419\tabularnewline
\hline 
R = 36  & 0.00347 & 0.00477\tabularnewline
\hline 
R = 72 & 0.00174 & 0.00318\tabularnewline
\hline 
\end{tabular}\caption{\label{table2}Change in z-coordinate for the intersection contour
for cylindrical surfaces}
\end{table}

\section{\label{sec:Conclusion}Conclusion}

In the present study, we have investigated the dynamics of flows past a stationary circular cylinder
embedded on spherical and cylindrical surfaces using a DEC-based numerical method at a modest Reynolds number of 100. Various surface curvatures were considered in order to explore the curvature effects on the flow dynamics, and key quantities such as drag, lift coefficients, the vortex shedding frequency, and the magnitude and spacing of the coherent vortices in the wake,  are compared with the flow past a cylinder on a flat surface. For a circular cylinder of unit diameter, the embedding spherical and cylindrical surfaces of radii ranging from 12 to 72 were considered. No significant differences were observed in the flow dynamics on the curved surfaces, when compared with the flow on a flat surface. We examined the magnitude of the curvature term in the momentum equation, relative to the other viscous term and found this to be insignificant. In addition, the cylinder contour geometry either is a planar circle as for the case of the embedding spherical surfaces case, or has very minor deviation from a planar circle in the case of the embedding cylindrical surfaces case. The deviation from a planar circle is small enough and the resulting effect on the flow dynamics is also insignificant.  A key conclusion is that the dynamics of a flow past a stationary circular cylinder embedded on a surface is not significantly affected by the curvature of the embedding surface, unlike what one may expect based on pure intuition.  The following open questions remain: what if the embedding surface curvature was even larger? and how will the results change with increasing Reynolds numbers. We leave exploration of these questions for the future. 

\bibliographystyle{apa}
\bibliography{manuscript}

\begin{thebibliography}{}

\bibitem[\protect\astroncite{Desbrun et~al.}{2005}]{desbrun2005discrete}
Desbrun, M., Hirani, A.~N., Leok, M., and Marsden, J.~E. (2005).
\newblock Discrete exterior calculus.
\newblock {\em arXiv preprint math/0508341}.

\bibitem[\protect\astroncite{Fage}{1929}]{fage1929effects}
Fage, A. (1929).
\newblock The effects of turbulence and surface roughness on the drag of a
  circular cylinder.
\newblock {\em Rep, Memo.}, 1283:1.

\bibitem[\protect\astroncite{Hirani}{2003}]{hirani2003discrete}
Hirani, A.~N. (2003).
\newblock {\em Discrete exterior calculus}.
\newblock PhD thesis, California Institute of Technology.

\bibitem[\protect\astroncite{Le et~al.}{2006}]{le2006immersed}
Le, D., Khoo, B.~C., and Peraire, J. (2006).
\newblock An immersed interface method for viscous incompressible flows
  involving rigid and flexible boundaries.
\newblock {\em Journal of Computational Physics}, 220(1):109--138.

\bibitem[\protect\astroncite{Li et~al.}{2009}]{Li2009}
Li, Y., Zhang, R., Shock, R., and Chen, H. (2009).
\newblock {Prediction of vortex shedding from a circular cylinder using a
  volumetric Lattice-Boltzmann boundary approach}.
\newblock {\em European Physical Journal: Special Topics}, 171(1):91--97.

\bibitem[\protect\astroncite{Mittal}{2005}]{Mittal2005}
Mittal, S. (2005).
\newblock {Excitation of shear layer instability in flow past a cylinder at low
  Reynolds number}.
\newblock {\em International Journal for Numerical Methods in Fluids},
  49(10):1147--1167.

\bibitem[\protect\astroncite{Mohamed et~al.}{2016}]{mohamed2016discrete}
Mohamed, M.~S., Hirani, A.~N., and Samtaney, R. (2016).
\newblock Discrete exterior calculus discretization of incompressible
  navier--stokes equations over surface simplicial meshes.
\newblock {\em Journal of Computational Physics}, 312:175--191.

\bibitem[\protect\astroncite{Mullen et~al.}{2009}]{mullen2009energy}
Mullen, P., Crane, K., Pavlov, D., Tong, Y., and Desbrun, M. (2009).
\newblock Energy-preserving integrators for fluid animation.
\newblock {\em ACM Trans. Graph.}, 28(3):38.

\bibitem[\protect\astroncite{Niemann and
  H{\"o}lscher}{1990}]{niemann1990review}
Niemann, H. and H{\"o}lscher, N. (1990).
\newblock A review of recent experiments on the flow past circular cylinders.
\newblock {\em Journal of Wind Engineering and Industrial Aerodynamics},
  33(1-2):197--209.

\bibitem[\protect\astroncite{Nitschke et~al.}{2017}]{nitschke2017discrete}
Nitschke, I., Reuther, S., and Voigt, A. (2017).
\newblock Discrete exterior calculus (dec) for the surface navier-stokes
  equation.
\newblock In {\em Transport Processes at Fluidic Interfaces}, pages 177--197.
  Springer.

\bibitem[\protect\astroncite{Park et~al.}{1998}]{Park1998}
Park, J., Kwon, K., and Choi, H. (1998).
\newblock {Numerical solutions of flow past a circular cylinder at Reynolds
  numbers up to 160}.
\newblock {\em KSME International Journal}, 12(6):1200--1205.

\bibitem[\protect\astroncite{Posdziech and Grundmann}{2001}]{Posdziech2001}
Posdziech, O. and Grundmann, R. (2001).
\newblock {Numerical simulation of the flow around an infinitely long circular
  cylinder in the transition regime}.
\newblock {\em Theoretical and Computational Fluid Dynamics}, 15(2):121--141.

\bibitem[\protect\astroncite{Qu et~al.}{2013}]{Qu2013}
Qu, L., Norberg, C., Davidson, L., Peng, S., and Wang, F. (2013).
\newblock {Quantitative numerical analysis of flow past a circular cylinder at
  Reynolds number between 50 and 200}.
\newblock {\em Journal of Fluids and Structures}, 39:347--370.

\bibitem[\protect\astroncite{Rao et~al.}{2015}]{rao2015review}
Rao, A., Radi, A., Leontini, J.~S., Thompson, M.~C., Sheridan, J., and
  Hourigan, K. (2015).
\newblock A review of rotating cylinder wake transitions.
\newblock {\em Journal of Fluids and Structures}, 53:2--14.

\bibitem[\protect\astroncite{Russell and Wang}{2003}]{russell2003cartesian}
Russell, D. and Wang, Z.~J. (2003).
\newblock A cartesian grid method for modeling multiple moving objects in 2d
  incompressible viscous flow.
\newblock {\em Journal of Computational Physics}, 191(1):177--205.

\bibitem[\protect\astroncite{Shi et~al.}{2004}]{Shi2004}
Shi, J.~M., Gerlach, D., Breuer, M., Biswas, G., and Durst, F. (2004).
\newblock {Heating effect on steady and unsteady horizontal laminar flow of air
  past a circular cylinder}.
\newblock {\em Physics of Fluids}, 16(12):4331--4345.

\bibitem[\protect\astroncite{St{\aa}lberg et~al.}{2006}]{Stalberg2006}
St{\aa}lberg, E., Br{\"{u}}ger, A., L{\"{o}}tstedt, P., Johansson, A.~V., and
  Henningson, D.~S. (2006).
\newblock {High order accurate solution of flow past a circular cylinder}.
\newblock {\em Journal of Scientific Computing}, 27(1-3):431--441.

\bibitem[\protect\astroncite{Taylor}{1915}]{taylor1915pressure}
Taylor, G.~J. (1915).
\newblock Pressure distribution on a cylinder.
\newblock {\em Techn. Rep. of Adv. Comm. for Aeron., London}, 1916:32.

\bibitem[\protect\astroncite{Thom}{1933}]{thom1933flow}
Thom, A. (1933).
\newblock The flow past circular cylinders at low speeds.
\newblock {\em Proceedings of the Royal Society of London. Series A, Containing
  Papers of a Mathematical and Physical Character}, 141(845):651--669.

\bibitem[\protect\astroncite{Von~K{\'{a}}rm{\'{a}}n}{1912}]{Karman1912}
Von~K{\'{a}}rm{\'{a}}n, T. (1912).
\newblock Uber den mechanismus des flussigkeits-und luftwiderstandes.
\newblock {\em Phys. Z.}, pages 49--59.

\bibitem[\protect\astroncite{Wieselsberger}{1921}]{wieselsberger1921neuere}
Wieselsberger, C. (1921).
\newblock Neuere feststellungen uber die gesetze des flussigkeits und
  luftwiderstands.
\newblock {\em Phys. Z.}, 22:321.

\end{thebibliography}

\end{document}